\documentclass[conference]{IEEEtran}
\usepackage{cite}
\usepackage{amsmath}
\usepackage{enumitem}
\usepackage{amssymb}
\usepackage{amsthm}
\usepackage{graphicx,epstopdf}
\usepackage{multicol}
\usepackage{textcomp}
\usepackage{multirow}
\usepackage{flushend}
\usepackage{color}
\makeatletter
\newcommand*\bigcdot{\mathpalette\bigcdot@{0.85}}
\newcommand*\bigcdot@[2]{\mathbin{\vcenter{\hbox{\scalebox{#2}{$\m@th#1\bullet$}}}}}
\makeatother
\usepackage{lipsum}
\usepackage{cuted}
\usepackage{wrapfig, blindtext}

\begin{document}
\title{Cybersecurity of Electric Vehicle Smart Charging Management Systems}


\author{Narayan Bhusal, \emph{Student Member, IEEE},  Mukesh Gautam, \emph{Student Member, IEEE}, \\and Mohammed Benidris, \emph{Member, IEEE}\\ 
Department of Electrical \& Biomedical Engineering, University of Nevada, Reno, NV\\ 
(emails: \{bhusalnarayan62,  mukesh.gautam\}@nevada.unr.edu, and mbenidris@unr.edu) \vspace{-1ex}}

\maketitle
\begin{abstract}
In concept, a smart charging management system (SCMS) optimizes the charging of plug-in vehicles (PEVs) and provides various grid services including voltage control, frequency regulation, peak shaving, renewable energy integration support, spinning reserve, and emergency demand response. These functionalities largely depend upon data collected from various entities such as PEVs, electric vehicle supply equipment (EVSE), service providers, and  utilities. SCMS can be susceptible to both cyber and physical threats (e.g. man-in-the-middle attack, data intrigued attack, denial of charging, physical-attack) due to interactions of and interdependencies between cyber and physical components. Cyber-physical threats through highly connected malware vectors  raise various concerns including public safety hazards to vehicle operators and those in the immediate vicinity as well as disruptions to electric grid operations. This paper describes the concept of SCMS and provides a comprehensive review of cybersecurity aspects of EVSEs and SCMSs with their possible impacts on the power grid and society. It also contributes to the development of cybersecurity measures to the SCMSs. Various functions of SCMS are reviewed in detail including peak shaving, demand charge reduction, frequency regulation, spinning reserve, renewable integration support, distribution congestion management, reactive power compensation, and emergency demand response with unidirectional PEVs charging. Also, a critical literature survey on current practices of SCMS cybersecurity is provided to explore major impacts and challenges of cyber-physical attacks and to identify research gaps and vulneraibilities in currently available SCMSs technologies.

\end{abstract}
\IEEEpeerreviewmaketitle

\section{INTRODUCTION}
Penetration of plug-in electric vehicles (PEVs) into power grid is proliferating worldwide. With this increased penetration, optimum management of PEV charging is becoming an important factor for PEV owners, service providers, utilities, and power system operation. For proper management of PEV loads, a smart charging management system (SCMS) that is capable to optimize charging of PEVs at both public charging and residential charging stations is necessary. Apart from shifting loads to more desirable times for the grid, the SCMS can contribute in grid services such as voltage and frequency support and seamless integration of renewable energy sources. To provide multiple services, SCMS should properly monitor and control PEVs that are connected to electric vehicle supply equipments (EVSEs) at both individual charging locations and congregated charging stations. In this context, monitoring and control of PEVs and EVSEs necessitate the real-time data communication with SCMS and the grid. Consequently, SCMS can be susceptible to several cyber-physical attack vectors. Therefore, the cyber-physical security of SCMS should be properly addressed to ensure that the system is hardened against attacks and be able to detect and mitigate attacks on the grid, charging networks, and customers in real-time.

Numerous studies have focused on the technical aspects of PEV integration and their impacts on the smart grid. Several technical aspects have been studied in \cite{PNNL201950,8911848,GARCIAVILLALOBOS2014717,TAN2016720,8705332} including charging strategies, energy management, power losses, grid interface technologies, renewable energy integration support, power system reliability, voltage and frequency regulation, and regulation of PEVs in electricity markets.

Although cybersecurity of the smart grid has been amply addressed in the literature, cybersecurity aspects of PEVs, EVSEs, and SCMS have not been fully addressed. In \cite{8887286}, a comprehensive review of the detection algorithm for false data injection attacks (FDIAs) in smart grids has been provided. FDIA detection algorithm has been divided into model-based (based on the dynamic models of systems) approach and data-driven (based on the utilization of available smart grid data) approaches. State-of-the-art approaches for FDIA against smart grids from different perspectives have been provided in \cite{7438916}. Also, \cite{7438916} explores the theoretical basis of FDIA, the physical and economic impacts of FDIA, basic defense strategies, and future research directions. Both \cite{8887286} and \cite{7438916} provide a rich source of references for further exploration of the cybersecurity of smart grids. Although the cybersecurity of SCMS possesses some similarities with that of smart grids, there are some fundamental differences. More research attention is needed to identify these differences and to develop cybersecurity measures for SCMS.

Due to interactions of and interdependencies between cyber and physical components of SCMSs, they can be susceptible to cyber-physical threats. Therefore, as a stepping stone toward the development of cybersecurity measures to SCMS, this paper describes the concept of SCMS and provides a comprehensive review of cybersecurity concerns of EVSE and SCMS along with possible mitigation measures. Various functions of SCMS are described in detail including peak shaving, demand charge reduction, frequency regulation, spinning reserve, renewable integration support, distribution congestion management, reactive power compensation, and emergency demand response. A survey of the existing literature on cybersecurity of SCMS is also provided. Furthermore, major cyber-physical attacks, impact of these attacks on SCMS, and major challenges to deal with these attacks are identified. Research gaps and vulnerabilities associated with current commercially available SCMS technologies are also highlighted.

The rest of the paper is organized as follows. Section \ref{scms} explains the concept and architecture of SCMS. Section \ref{grid_services} explores potential grid services that PEVs can provide via SCMS. Section \ref{Existing_work} reviews existing literature on cybersecurity of PEVs, EVSEs, and SCMSs. Section \ref{cyber-physical-attacks} discusses the major cyber-physical threats in SCMS. Section \ref{research_gap} presents the research gaps and vulnerabilities associated with commercially available SCMS technologies. Finally section \ref{conclusion} provides concluding remarks. 

\section{Concept and Architecture of Smart Charging Management System}\label{scms}
\begin{figure}
    \centering
    \includegraphics[scale=0.35]{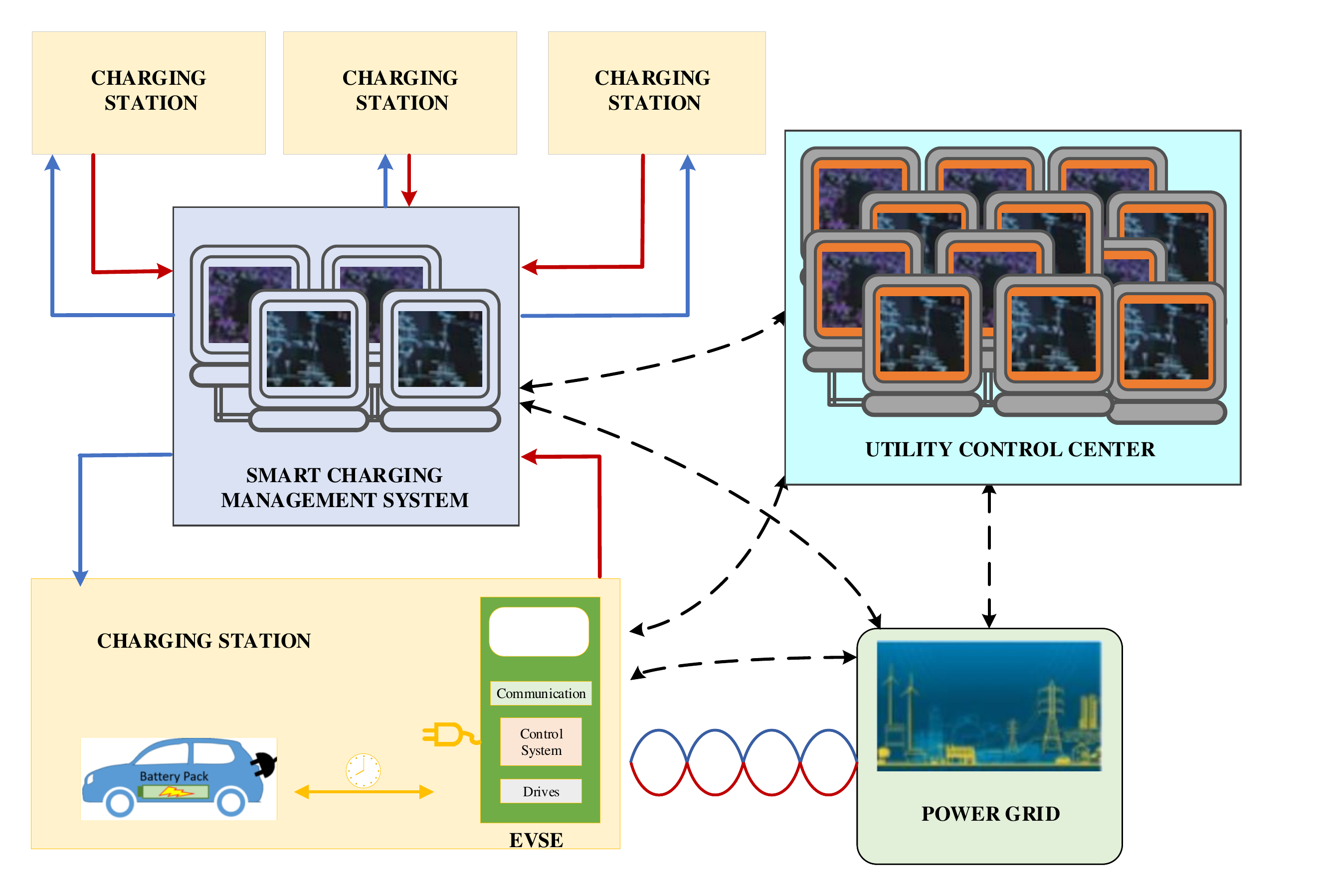}
    \caption{Architecture of Smart Charging Management System}
    \label{fig:SCMS}
\end{figure}

SCMS is a system that provides various benefits to PEV owners, charging network operators, energy service providers, and  flexibility to the electric grid operators through the utilization of flexible PEV loads. 
Apart from optimizing the charging of PEV loads, SCMS also provides various grid services such as peak shaving, voltage control, frequency regulation, renewable integration support, demand-side management, demand charge reduction, loss reduction, and emergency demand response, to name a few. Fig.~\ref{fig:SCMS} provides a typical architecture for SCMS to describe its concept of operation \cite{PNNL201950, GARCIAVILLALOBOS2014717}.

We can see from Fig.~\ref{fig:SCMS} that SCMS communicates with charging stations and its EVSEs, electric power grid, and electric utility to manage the charging of PEVs and to provide various grid services. The charging stations in the figure could be individual charging station or congregated charging stations at workplace, multi-unit dwelling, and retail-establishment. Each charging station is equipped with one or more EVSEs that can communicate with SCMS and the power grid for controlling the charging process and to provide various real-time measurements. SCMS receives charging requests from PEVs and EVSEs and electricity price signals and various grid service requests from utility control centers. The real-time and historical data from EVSE are processed and analyzed at SCMS and utilized to make demand forecast, planning, and investment decisions. Section \ref{grid_services} provides a brief description of potential grid services that PEVs can provide via SCMS. Also, providing cyber-physical security measures to detect, identify, control, and mitigate these threats is an important part of SCMSs.

\section{Potential Grid Services Provided by PEVs via SCMS}\label{grid_services}
There are several challenges and complications associated with vehicle to grid (V2G) integration \cite{TAN2016720}. Apart from that, there is still a big question on the economic feasibility of V2G integration \cite{GTM1932018, GOUGH201712, INL201298}. In other words, V2G technology is not something that will come into practice anytime soon. Therefore, in this paper, only grid services that are possible from the flexibility of charging PEV loads at different times and rates are considered. In other words, this paper describes how PEV loads can be utilized to provide peak shaving, reactive power compensation, frequency regulation, renewable energy integration support, etc. through controlling charging time and rates without the use of V2G service. This section describes in detail the grid services provided by SCMS utilizing PEV loads.

\subsection{Peak Shaving}
Peak shaving refers to the reduction of peak power demand of the power grid. Peak shaving can be achieved by controlled charging through adjusting PEV charging times and rates. Very few or no PEVs are charged (or charged at a very low rate) during the time of peak power demand---charging of low-priority PEVs can be shifted to off-peak hours. Peak shaving can be implemented using SCMS algorithms. SCMS uses different pricing schemes for different times of the day (imposing high prices to charge the PEVs during the period of high power demand and imposing low prices during the period of low power demand). Numerous literature have studied the peak shaving scheme using PEVs \cite{IOAKIMIDIS2018148, 10.1115/1.4031209, 6465656, TAN2016720, 5969519}.

\subsection{Ancillary Service}
\textit{Frequency regulation:} frequency drop usually occurs when the generation deficit occurs in the power system. Electric utilities constantly balance the supply and demand to keep system frequency at its nominal value. When a large number of PEVs are charged in an uncoordinated manner, it can disrupt the balance between generation and consumption resulting in disruption in system frequency. Therefore, providing the balance between power consumption and generation through flexible loads is critical if the generation cannot ramp up/down within a short period of time. The direct real-time control of the charging of PEVs can provide the power balance and support to regulate the frequency. The charging rate of PEVs is regulated up and down to meet the preference operation point to keep the frequency within the desired range \cite{LOPEZ2015689}.

\textit{Spinning reserve:} this service can be achieved using PEVs by lowering the charging rate during the time of higher power demand \cite{6084772}. A proper compensation scheme should be provided to PEVs for being agreed to provide these services.

\subsection{Support Renewable Energy Integration}
The ability to support renewable energy integration into the electric power grid is one of important transformative impacts of PEVs. Charging PEVs at higher power production mitigates the problem of photovoltaic/wind generator over-generation. The SCMS performs this by managing to charge a large number of PEV loads when there is peak power production from renewable energy sources (e.g. photovoltaic and wind). The integration of renewable sources can be achieved via real-time electricity pricing and demand response programs\cite{5497937}. Because of the coordination between charging of PEVs and renewable generation, charging of PEVs and the excess generation of renewable sources do not need to be curtailed. Several approaches have been presented to demonstrate the ways of supporting renewable integration via PEVs \cite{5399955, MWASILU2014501, RICHARDSON2013247}.

\subsection{Reactive Power Compensation}
It is highly essential to understand the reactive power behavior and capability of PEVs for reactive power compensation. The reactive power behavior of PEVs has been studied in \cite{6672752} which has shown that all of the studied PEVs have a power factor above $0.99$. This indicates that the reactive power support can be provided by injecting reactive power during the time of charging (i.e. setting PEV's charger to provide capacitive power factor) \cite{LEEMPUT201524, 7339484 }. However, this comes with the cost of increased apparent power rating resulting in more active power consumption. For example, to obtain $0.95$ capacitive power factor, the apparent power rating of a PEV becomes $105.3 \%(1/0.95)$. This $5.3\%$ increase in PEV charger can cause up to $\sqrt{1.053^2-1}=32.9\%$ increase in active power consumption. For more work on reactive power supporting capability of PEVs, refer \cite{8274266,7339484,LEEMPUT201524,6779608}.

\subsection{Emergency Demand Response}
Emergency demand response is implemented in the utility control center during unexpected contingencies. These contingencies may disrupt the balance between generation and consumption. As PEVs are flexible loads, based on the command signal from the control center, they can be stopped from charging in response to emergency demand response \cite{CEC2018100, 5484613}.

\section{Literature Survey on Cybersecurity of SCMS}\label{Existing_work}
Cyber-physical threats such as man-in-the-middle attack, data integrity attack, payment fraud, privacy/tracking concerns, intentional charging or discharging, denial of service attack, malware injection with the help of PEVs, and rapid cycling of a large number of PEVs are potential attacks on SCMS and its integrated components \cite{DOEDOT201844}. These cyber-physical threats result in various consequences: public safety hazard to the vehicle operators and those in the immediate vicinity and initiating and exacerbating electric grid disruption. Although a considerable amount of work has been proposed for the cybersecurity of the smart grid,  cybersecurity of SCMS has received very little attention. Some of the literature that make an important contribution toward the cybersecurity of PEVs are as follows. 

Authors of \cite{8960519} have proposed a control-oriented approach to detect cyber-attacks that can affect PEV batteries during charging. Two algorithms have been proposed to detect cyber-attacks during the charging as follows:  (a) static detector utilizing measured variables and (b) dynamic detector which utilizes the dynamics as well as the measurement variables. The results demonstrated in this study show that the performance of the dynamic detector is better than the static detector.

Authors of \cite{Abedi2015} have explored the cybersecurity measures of PEVs in the smart grid with a review on some of the state-of-the-art algorithms that have been used to detect cyber-attacks. The impact on operational cost, net power demand, and charging/discharging algorithms of data attacks on a PEV smart parking lot has been studied. Two intrusion detection approaches (model-based and signal-based with specific application to PEV data attack) have also been examined in \cite{Abedi2015}.

In \cite{8790593}, vulnerability analysis and risk assessment of cyber-physical attacks on PEVs charging networks have been described. In \cite{8767141}, a cyber-physical interaction of various components of EVSE is provided with the classification of associated vulnerabilities and cyber-physical threats. Cybersecurity of the battery management system of PEVs has been considered in \cite{8440144} through a neural network-based approach that estimates state of charge (SOC) of a battery under cyber-attacks. However, the work in \cite{8440144} does not address the cybersecurity concern from the perspective of the charging management system. Although the aforementioned literature studied some insights into the cybersecurity-related concerns, there are still several areas of charging management systems that are unexplored; therefore, further research is needed. 

\section{Major Cyber-physical Attacks in SCMS}\label{cyber-physical-attacks}
SCMSs and EVSEs are connected to PEVs, building energy management systems, electricity grid, telecommunication networks, billing systems, and utility control centers. These can cause disruption of electricity load management for buildings or the electric grid or damage to PEV batteries. The accessibility and power consumption of an EVSE system is a potential mechanism for disrupting the power of a building or distributing electricity to a specific area. In addition, if a hacker installs persistent malware in EVSE and propagates it to SCMS and power grid, it will exacerbate the disruption even further. In this context, SCMS and its integrated network will not be able to meet confidentiality, integrity, and availability (CIA) requirement. The major cyber-physical attacks in SCMS and its associated network are described as follows.

\subsection{False Data Injection Attack}
PEV charging/discharging data throughout the grid are collected with the help of smart measuring devices connected to the EVSEs, installed at charging stations. All of these grid services that SCMS provides depend upon the charging/discharging request received from PEVs and measurements provided by EVSEs. Data collected from SCMS and its integrated system are analyzed by utilities to determine the optimal dispatch of generators, demand-side management, demand forecast, and reliability and stability analysis of the system, to name a few. As PEV charging/discharging data have large use, their accurate measurement, processing, and analysis are very important. FDIA is one of the crucial attacks that can cause several damages to the PEVs, EVSE, SCMS, and even to the grid. FDIA aims at manipulating SCMS and its integrated system-related various data such as \cite{Abedi2015}: (a) energy request, (b) energy usage, (c) price signal from a utility, (d) demand response bidding from EVSE, (e) demand response needs from the utility, (f) event messages, (g) PEV ID, (h) premise location ID, (i) utility ID, and (j) customer ID, communicated between PEVs, EVSE, and SCMS. FDIA can cause overcharging to batteries and several damages to PEVs and the grid.

Although little attention has been given to detect data attacks on SCMS, the various techniques that have been applied to identify FDIA in power systems could motivate the research on detection of FDIA in SCMS. Authors of \cite{8887286, 7438916} have provided a comprehensive review of FDIA and their detection from various perspectives for the smart grid. Both \cite{8887286} and \cite{7438916} provide a rich source of references for further exploration of the cybersecurity of smart grids. References \cite{6897944, 6982207, 6503599, WANG2019208, 7926429, SHUKLA2020100167} provide resources to develop appropriate cybersecurity measures for SCMS.

\subsection{Man-in-the-Middle Attack}
\begin{figure}
\hspace{-1ex}
    \includegraphics[scale=0.57]{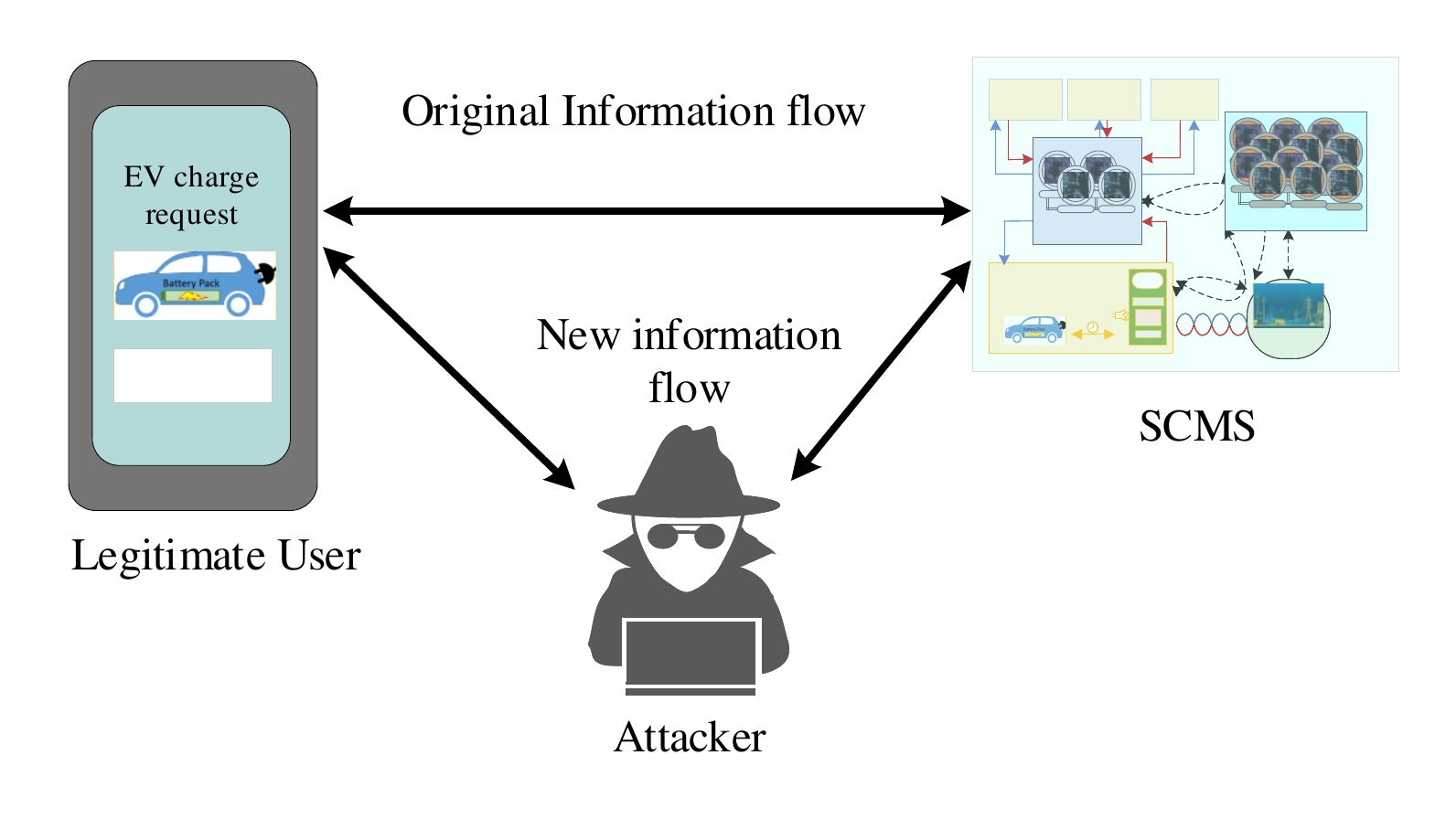}
    \caption{Basic structure of Man in the middle attack}
    \label{fig:man_in_middle}
\end{figure}

In Man-in-the-Middle (MITM) attack, an attacker intercepts and manipulates data that are communicated between various parties \cite{8366503}. Fig.~\ref{fig:man_in_middle} shows a basic block diagram of MITM attack. In the context of SCMS, an attacker can intercept communication between PEVs, EVSEs, and SCMS and modify, drop, and falsify data transmission. When attackers insert between PEVs, EVSEs, and SCMSs, they can create tracking issues, payment fraud (e.g. the charging cycle does not last full amount of time paid for, the charger is spoofed into providing free service), and violate other personal privacy \cite{DOEDOT201844}. Through MITM, an attacker can also cause intentional overcharging/discharging of PEV batteries causing damage to PEV and its batteries and taking the PEVs out of service or degrading range. MITM attacks can also overload distribution transformers and sometime power grid frequency and voltage stability disturbances leading to power grid failure via rapid cycling of a large number of PEV loads \cite{INL201813}. Apart from that, compromised EVSEs and PEVs can cause several personal safety concerns.

Authors of \cite{8960519} have provided some studies on impacts of overcharging attacks on PEV batteries; however, there is still a huge research gap for MITM attacks in SCMSs. Work proposed in \cite{10.1145/1023646.1023662, SU2018122, doi:10.1177/1550147717741463, 4768661, 8344724, 10.1007/978-3-540-89255-7_8, 4683982} could be adopted to develop detection, identification, defense, and mitigation measures for MITM attacks in SCMSs.

\subsection{Denial-of-Service Attack}
A denial-of-service attack causes the unavailability of network service to intended users as a result of an attacker's action to jam and overload the network \cite{8049303}. In the case of SCMS and its entities, attackers can attack servers and block valid requests from PEVs resulting in rejecting requests from legitimate PEV users. Due to denial-of-charge (a type of  DoS in SCMSs), important emergency vehicles (e.g., ambulance, firetrucks, and security vehicles) may be denied from charging resulting in the detrimental effects on the various emergency services and society. 

Although very few works have been proposed for DoS attack in SCMS, authors of \cite{8960519} have provided some insightful studies on the denial-of-charging attacks in which measurements based static attack detection and measurements and system dynamics based dynamic attack detection algorithms have been proposed. The work proposed in \cite{8630919, 8692706, DCIS-1012/18} could be used as a basis to develop DoS detection, defense, and mitigation system for SCMSs. 

\subsection{Malware Injection via EVSEs}
Due to the publicly available nature of EVSEs, especially, at public charging stations, EVSEs are susceptible to malware injections. The malware injected EVSEs can cause theft of several sensitive information such as payment information (debit/credit card information), personal information, charging time, payment amounts, etc. \cite{INL201813}. The malware injected EVSE not only affects individual EVSEs, but it also has a probability to propagate to a network of EVSEs.  The malware injected in EVSEs can also pass to PEVs, SCMS, and the power grids resulting in detrimental effect to all the stakeholders\cite{DOEDOT201844}. The best way to deal with these attacks is to provide cybersecurity-related testing and assessment while installing EVSEs.

\subsection{Physical Attack}
Physical attacks to EVSEs or PEVs can compromise the service provided by EVSEs and PEVs. A compromised PEV or EVSE is a potential personal safety concern and grid network concern. The coordinated charging events could cause widespread disruption of the power grid\cite{INL201813}.

\section{Research Gaps and Associated Vulnerabilities of Commercially Available SCMS Technologies}\label{research_gap}
This section describes the research gaps and associated vulnerabilities of currently available technologies of various entities of SCMSs such as PEVs, EVSEs, and smart meters. The research gaps and vulnerabilities identified by DOE/DHS/DOT technical meeting on electric vehicle and charging station cybersecurity  are listed as follows \cite{DOEDOT201844}.
\begin{itemize}
     \item Currently available PEV and EVSE charging infrastructures are immature for cybersecurity best practices. Most of the PEV industries do not have security software and development methodologies and guidelines. Also, buyers of PEVs and EVSEs do not typically specify the cybersecurity-related protection requirements because of limited knowledge.
    \item The trust model for end-to-end communication is in an early stage of development. Also, the standards for end-to-end communications between PEVs, EVSEs, and the power grid are still in the development phase.   
    \item  Cybersecurity-related testing and assessment are not accessible to most of the PEVs and charging infrastructure industries. Further research in this field is inevitable. 
    \item The guidelines and guidance on cybersecurity requirements for wireless charging infrastructures for light passenger PEVs, electric buses, and electric trucks are still in the testing and demonstration phase.
    \item Currently available PEV infrastructures such as EVSEs, smart meters, advanced metering infrastructure, and demand response equipment are yet to be matured with up-to-date technologies. 
    \item Commonly available EVSEs are still struggling with proper physical security guidelines and guidance. Unavailability of such guidelines has adversely affected the consumer's confidence in PEVs. 
\end{itemize}



\section{Conclusion}\label{conclusion}
This paper has described the concept of SCMS and the potential grid services that can be provided by unidirectional PEVs (without the use of V2G technology) charging. The various grid services provided by SCMS such as peak shaving, demand charge reduction, frequency regulation, spinning reserve, renewable energy integration support, distributed congestion management, reactive power compensation, and emergency demand response were discussed in detail. This paper also provided a literature survey on existing work on cybersecurity of SCMS, the major cyber-physical attacks in SCMS with their various impacts, and the major challenges of dealing with these attacks were explored. Moreover, research gaps and associated vulnerabilities of commercially available SCMSs technologies were also discussed.

The goal of SCMS is not only to optimize the coordinated charging of a large number of PEVs but also provide numerous grid services. Therefore, proper algorithms should be developed through more research and development in SCMSs. Due to the involvement of various cyber-physical components during the implementation of SCMS, it possesses various serious cyber-physical threats. Therefore, proper detection, identification, defense, and mitigation measures are required to provide cybersecurity of SCMS. Moreover, current commercially available SCMS technologies are struggling with unique cybersecurity-related threats, therefore, further research and development, and various guidelines and security need to be developed.

\section*{Acknowledgement}
This work was partly supported by the U.S. National Science Foundation (NSF) under Grant NSF 1847578.


\bibliographystyle{IEEEtran}
\bibliography{References.bib}
\end{document}